\documentclass[10pt,dvips]{article}
\usepackage{epsfig,times}

\begin{document}

\begin{center}
{\bf An anti-knee} \\ [2mm]
{\it Igor Lebedev} \\ [2mm]
{Institute of Physics and Technology, Almaty, Kazakhstan}
\end{center}

\begin{abstract}
The energy spectra for five elemental groups (p, He, C, Si, Fe) of cosmic rays, are simulated. 
Calculations are based on an analysis of superposition of few cosmic ray sources (Supernovae 
of different types) and principles of diffusive shock acceleration. Results of the analysis in the 
knee region show that summarizing energy spectra for elemental mass groups of cosmic rays can 
have significant oscillations of spectral index $\gamma$. Moreover, a kink (increasing $\gamma$) 
in Fe spectrum at energy $2\cdot 10^{15}$ eV, is discovered. In order to distinguish from the knee 
(decreasing $\gamma$) the kink can be called an anti-knee (increasing $\gamma$). 
\end{abstract}

\section{Introduction}

The energy spectrum of cosmic rays (CR) follows a simple power law $dN/dE\sim E^\gamma$. The observation of the change of the spectral index $\gamma$ at energy of few PeV (the so-called the 'knee' \cite{1}) has induced considerable interest. Nevertheless, despite of about 50 years of experimental and theoretical activities, the origin of the knee phenomenon has not yet been convincingly explained.
 
Theoretical models explaining the knee can be divided on two categories. 
The first is based on astrophysical reasons, i.e. the knee is a property of energy spectrum. Second class considers new particle physics processes in the atmosphere. First category is developed more intensively. Different authors explain the knee as result of an acceleration process \cite{3}, propagation of cosmic ray particles through the Galaxy \cite{4}, interaction of cosmic rays with background particles in the Galaxy \cite{5}. The most of authors predict detailed spectra of CR elemental groups.  

Thus for a solution of the knee problem an accurate measurements of the energy spectra of individual cosmic ray elements (or elemental groups, at least) are necessary.
 
Unfortunately, above 1 PeV cosmic rays are experimentally accessible in ground-based installation only. Considerable fluctuations of showers in the atmosphere complicate the task of energy and mass reconstruction. 
A solution of the problem can be found by using correlation approaches and methods, which allow to reduce the influence on intrinsic fluctuations at shower development for the reconstruction of energy and mass of primary cosmic rays \cite{6b}-\cite{6l}.    

One of the most interesting recent results (from two-dimensional $N_e-N_\mu$-analysis of KASCADE data) in knee region is a break (a significant increase) of the slope of heavy elements spectrum at energy of few PeV \cite{6a}- \cite{6a2005}.    
 These results are presented in Fig.1. At that, it should notice that spectra of elemental groups in Fig.1left (QGSJET model) and in Fig.1right (Sibyll model) are different. Therefore, a behavior of these spectra can be discussed only (at least, in present stage). 

Nevertheless, the increasing $\gamma$ is important criterion for the most of theoretical models explaining the knee. 
Many of these models (see reviews \cite{horandel}-\cite{ptuskin}) predict unchangeable energy spectra of elemental groups 
with $\gamma$ from $\sim -2.6$ to $\sim -2.8$ at different assumptions. Such value of $\gamma$ must be kept before defined 
energy (cut off energy of accelerated particles). After that, a considerable decreasing $\gamma$ (the knee in spectra of CR 
elemental group) must be observed. The significant increase of the spectral index in PeV region is not predicted. In order to 
distinguish from the knee (decreasing $\gamma$) the kink (observed by KASCADE) can be called an ankle or an anti-knee (increasing $\gamma$). 
 
\begin{figure}[t]
\begin{center}
\includegraphics*[width=0.48\textwidth,angle=0,clip]{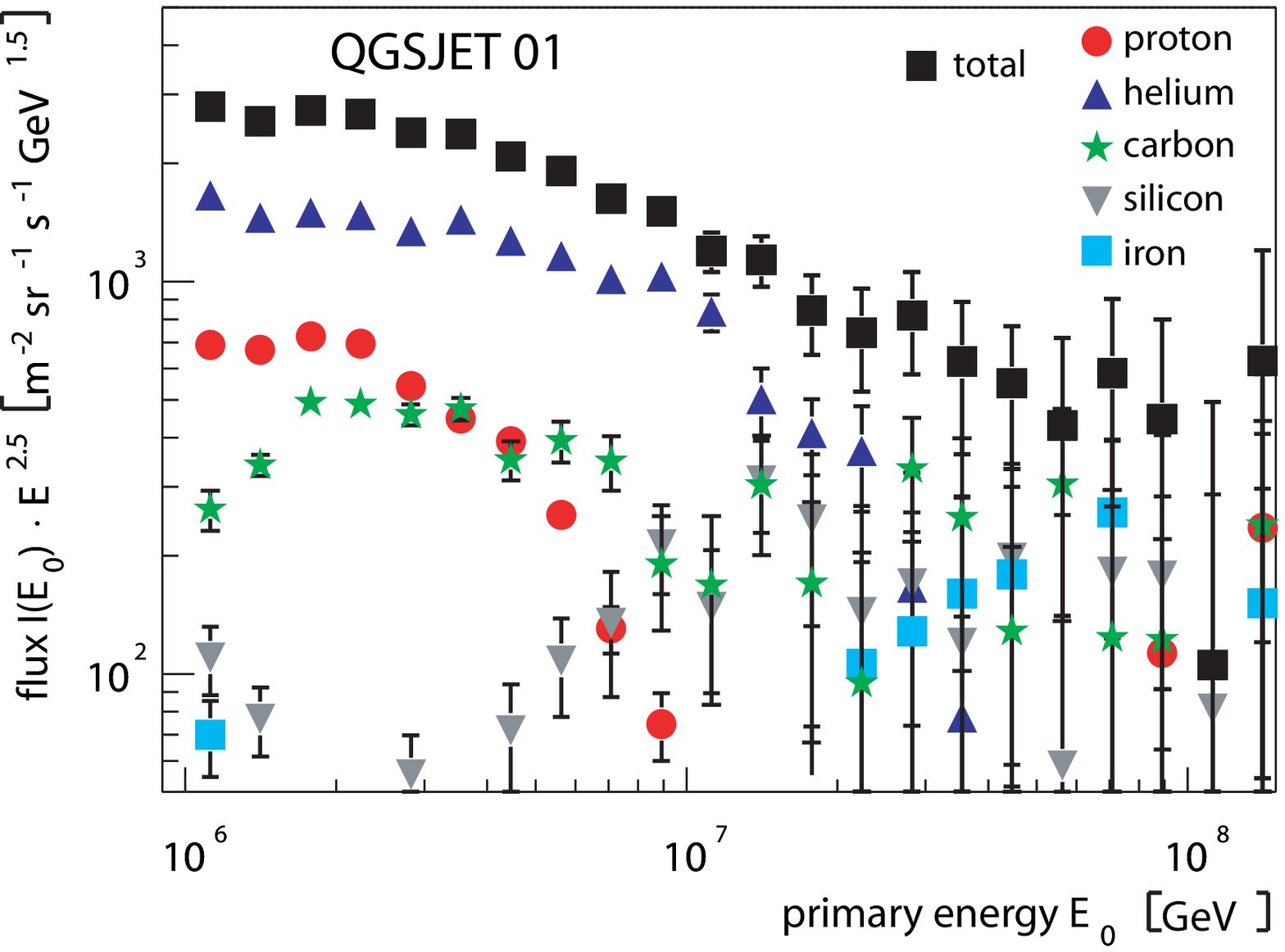}
\includegraphics*[width=0.48\textwidth,angle=0,clip]{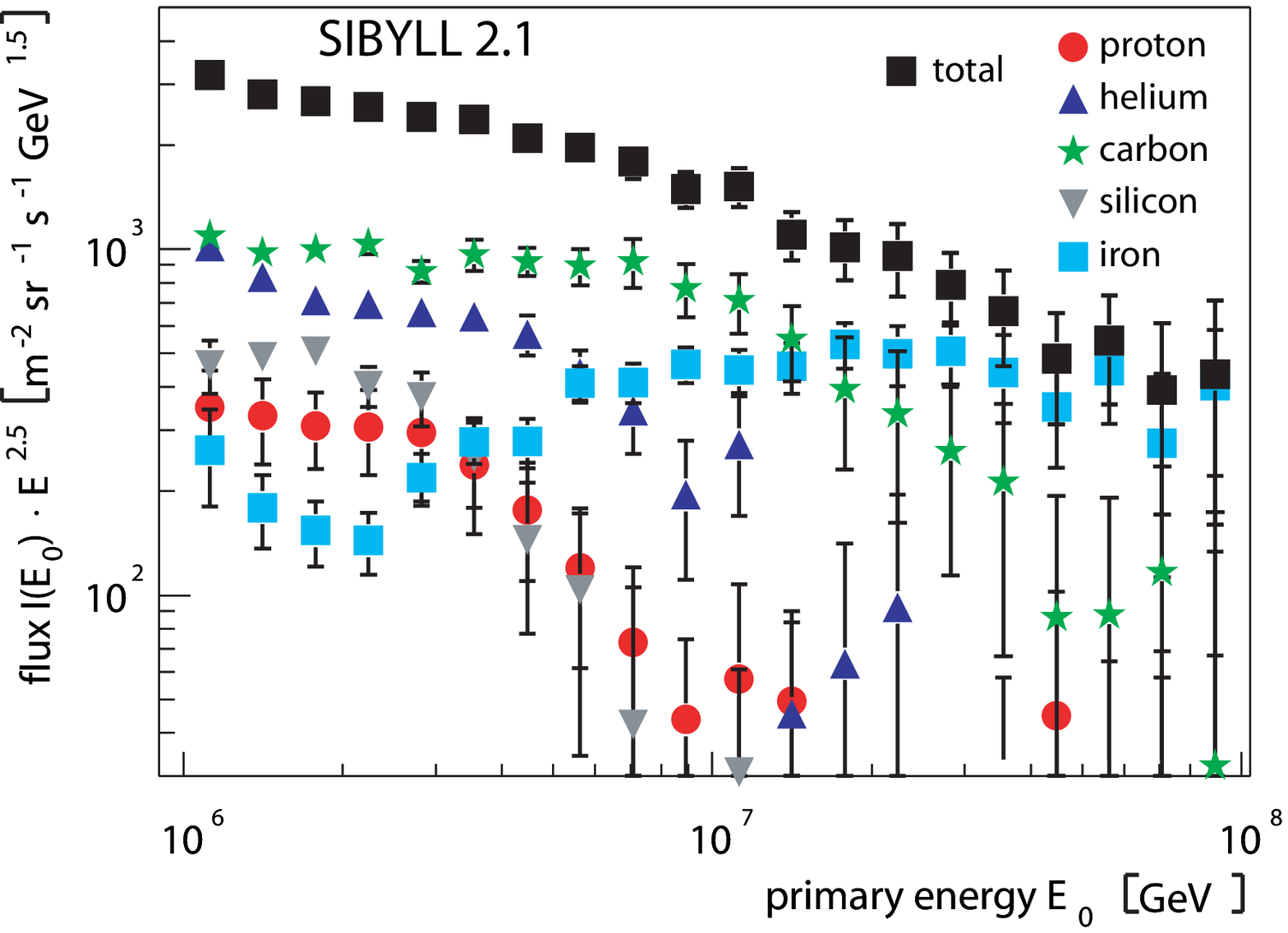}
\caption{\label{fig1} 
CR energy spectra of KASCADE \cite{6a}.
}
\end{center}
\end{figure}

In Fig.1right the anti-knee is observed for Fe spectrum at energy $2\cdot 10^{15}$ eV. In this region the spectral index increases 
from the value of $\gamma\sim -3$ (at least, $\gamma < -2.5$ before $2\cdot 10^{15}$ eV) to $\gamma\sim -2$ (at least, $\gamma>-2.5$ after $2\cdot 10^{15}$ eV). 
In Fig.1left Fe spectrum in this region isn't shown. But Si spectrum has similar peculiarity. 

Such behavior is not agreeing with predictions of the most of theoretical models. 
If these observations could be confirmed by excluding all experimental or reconstruction effects they would be very exciting.  

That the feature of the heavy elements spectrum is not totally impossible in sense of astrophysical scenarios will be shown below. 

In paper \cite{erlykin} it is discussed how a single supernova source can explain the KASCADE data.    

In present work, spectra of 5 elemental groups (p, He, C, Si, Fe) are simulated on the base of a superposition of few CR sources (Supernovae of different types). It allows to describe the behavior of energy spectra, observed by KASCADE.        

Supernova (SN) explosion marks the end of the life of some stars. The process produces extending shell of ejected matter. The diffusive propagation of high-energy particles allows them to cross the shock front many times and it allows to increase the energy of the particles.   
 
The model predicts \cite{7b1}-\cite{7a}:\\
1. The power-like and approximately similar spectra of various nuclei of cosmic rays: $dN/dE\sim E^{\gamma} $.\\
2. The slope of power-like source spectrum is $\gamma_{sour} \sim -2$.\\
3. Cut off energy of accelerated particles 
$E_{c}$ $\sim$ $Z$, where $Z$ is charge of nucleus. 
In the interval of $E>E_{c}$ the index $\gamma \sim -5$.

Hence, this standard model predicts  $\gamma_{sour} \sim -2$. But experimental observations give $\gamma_{obs} \sim -2.75$. For a solution of the problem the most of authors use some mechanisms for correction of $\gamma$, i.e. $\gamma_{obs} = \gamma_{sour} + \Delta\gamma$. The value of $\Delta\gamma=-0.6$ for simple diffusion model, $\Delta\gamma=-0.3$ for model with reacceleration, $\Delta\gamma=-0.8$ for model with Galactic wind. So the value of $\Delta\gamma$ is not well known yet. 

In present work the transition of $\gamma_{sour}\rightarrow \gamma_{obs}$ is simulated by a superposition of few cosmic ray sources with defined mass composition. 

It is possible few effects (superposition of different sources, propagation peculiarities, etc.) give a deposit to $\Delta\gamma$. Therefore, they should be combined at some stage. 
 
\section{Hypothesis}

Let's consider only one hypothetic CR source with the definite mass composition (p-46.7\%, He-28.3\%, C-13.5\%, Si-6.9\%, Fe-4.6\%). Such choice of mass composition provides $\gamma_{obs}\sim -2.75$ instead of $\gamma_{sour} = -2$ for a long energy interval: from proton $E_{c}$ to iron $E_{c}$.
The composition of p-54.2\%, He-32.5\%, C-7.6\%, Si-3.8\%, Fe-1.9\% brings $\gamma_{obs}\sim -3.1$. 

For a simulation of the all-particle spectrum in more wide energy interval (including the knee region) few CR sources with different $E_{c}$ are necessary.   
At that, a superposition of few sources complicates the problem significantly. 
In order to quote the indices of the individual component below and above the knee it needs to know $E_{c}$ of CR sources and/or CR mass composition at different energy.  


%
\begin{figure}[t]

\begin{center}

\includegraphics*[width=0.48\textwidth,angle=0,clip]{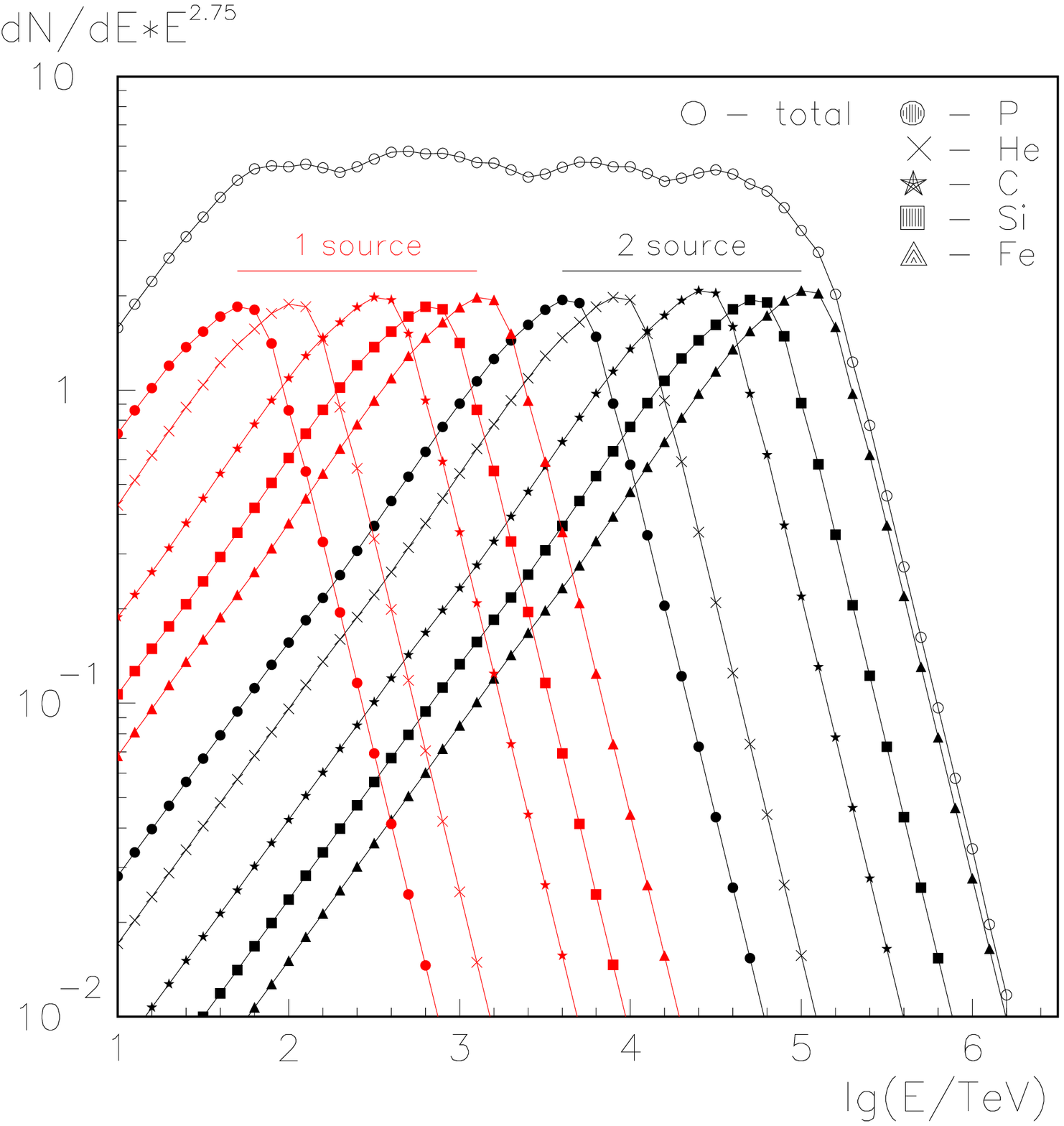}
\includegraphics*[width=0.48\textwidth,angle=0,clip]{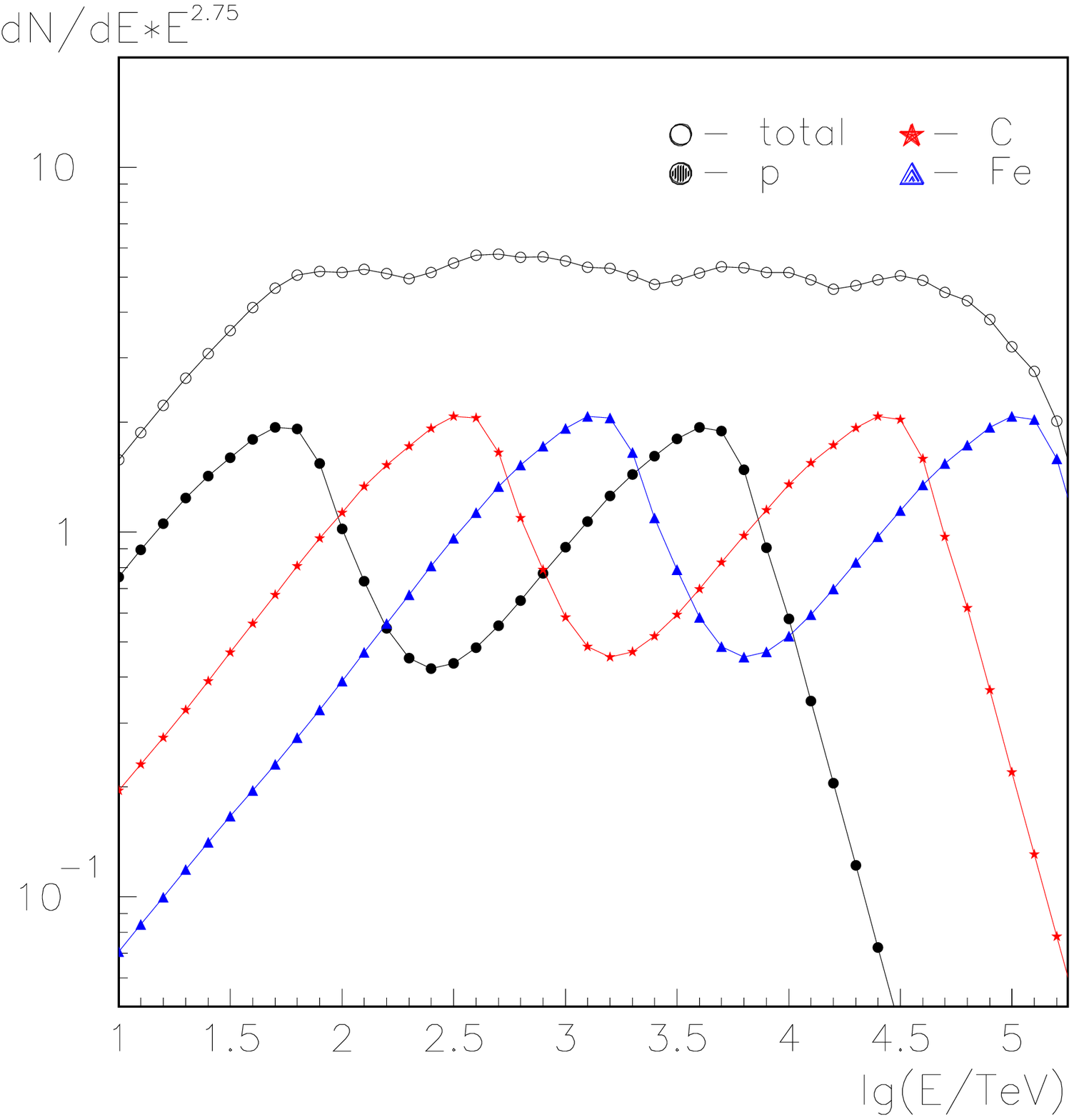}
\caption{\label{fig2} Left: Energy spectra (in relative units) for two hypothetic CR sources with the same mass composition: $p-46.7\%$, $He-28.3\%$, $C-13.5\%$, $Si-6.9\%$, $Fe-4.6\%$. 
Right: Summarized spectra from two sources.}
\end{center}

\end{figure}

Unfortunately, extensive air shower experiments cannot give one-valued inference about cosmic ray mass composition. At first, it is connected with low accuracy of mass reconstruction by EAS measurements. Second, the mass estimation depends on a choice of used hadronic interaction model. Nevertheless, some qualitative predictions can be useful for experimental search.
 
Let's consider two hypothetic cosmic ray sources with the same mass composition ($p-46.7\%$, $He-28.3\%$, $C-13.5\%$, $Si-6.9\%$, $Fe-4.6\%$). Moreover, let's choose a case, when the cut off energy of first source ($E_{c1}$) is much less the cut off energy of second source ($E_{c2}$). 

Spectra of five elemental groups for these hypothetic CR sources are presented in Fig2left. In this case, almost unchangeable slope of all-particle spectrum ($\gamma\sim -2.75$) in wide energy interval (from proton $E_{c1}$ to iron $E_{c2}$), is observed.

If the cut-off energy of second source would be lower (and therefore the difference between $E_{c1}$ and $E_{c2}$ would be less), a bump in all-particle spectrum would appear in the energy range of the overlap region. In this case, almost unchangeable all-particle spectrum can be simulated by a decrease of the Fe quote of first source and/or proton quote of second source. If the cut-off energy of second source were at higher energies, a dip in all-particle spectrum would occur. 
In given case it needs to increase the Fe quote of first source and/or the proton quote of second source. A superposition of few sources with $E_{c}$ of different types of Supernovae will be considered in section 3.

In Fig.2right the summarized elemental groups spectra from superposition of two sources, is presented. From the Figure it can be seen that the spectral index shows gigantic oscillations. $\gamma$ changes from $\sim -2$ to $\sim -5$ (a  knee) at energy of $E_{c1}$ and $E_{c2}$. In energy region between $E_{c1}$ and $E_{c2}$ the spectral index increases from $\gamma\sim -5$ to $\gamma\sim -2$ (an anti-knee), when deposit of second source in summarized spectrum becomes significant. 

In Fig.3left the mean logarithmic mass from superposition of two CR sources, is presented.
At that the mean logarithmic mass increases from proton $E_{c}$ to iron $E_{c}$. In energy interval of between iron $E_{c1}$ and proton $E_{c2}$ the mean logarithmic mass decreases considerably.

Therefore, two sources with large difference of $E_{c}$ can initiate significant changes of the mean logarithmic mass and the summarized energy spectrum of elemental groups.

\begin{figure}[t]
\begin{center}
\includegraphics*[width=0.48\textwidth,angle=0,clip]{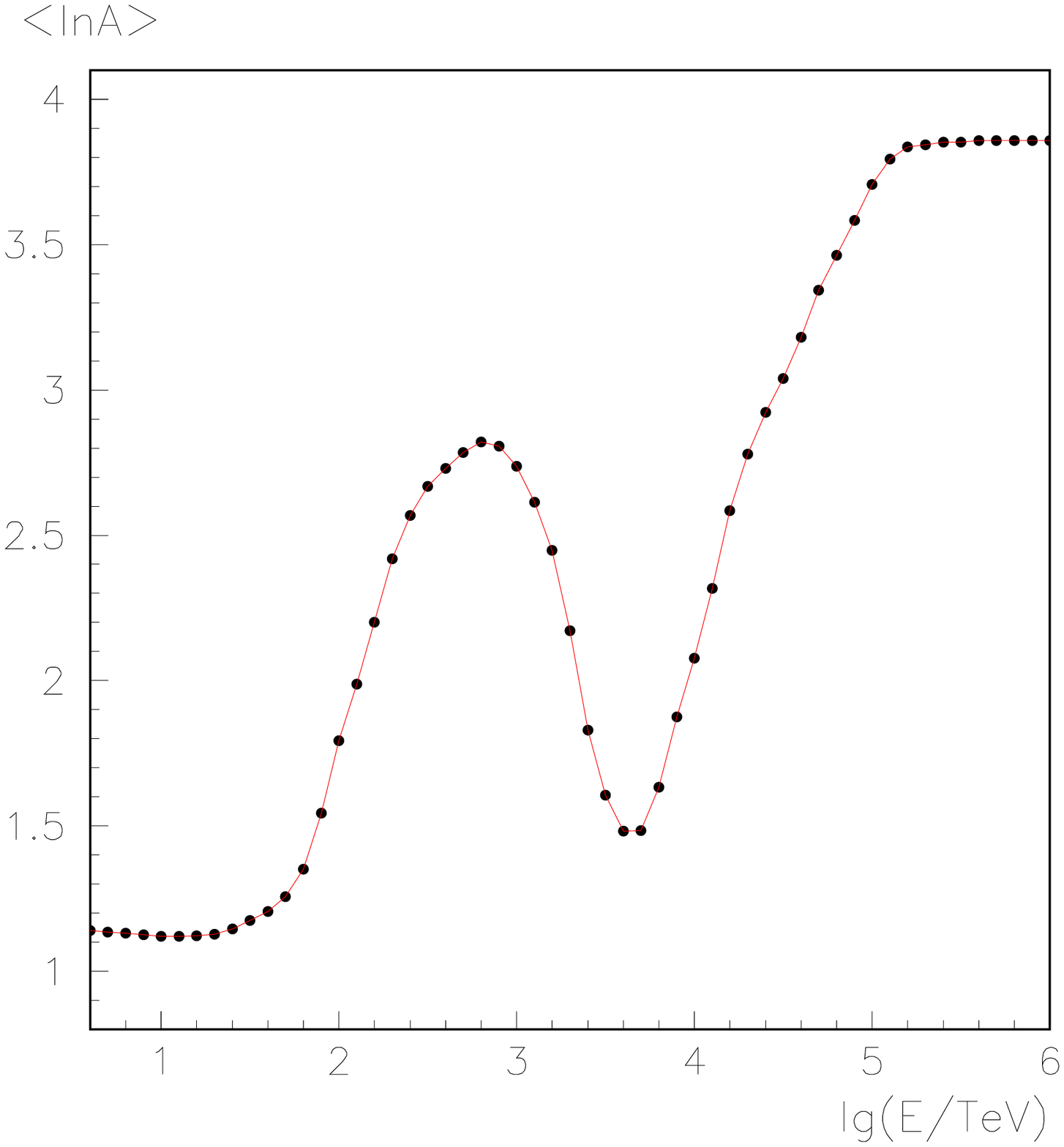}
\includegraphics*[width=0.48\textwidth,angle=0,clip]{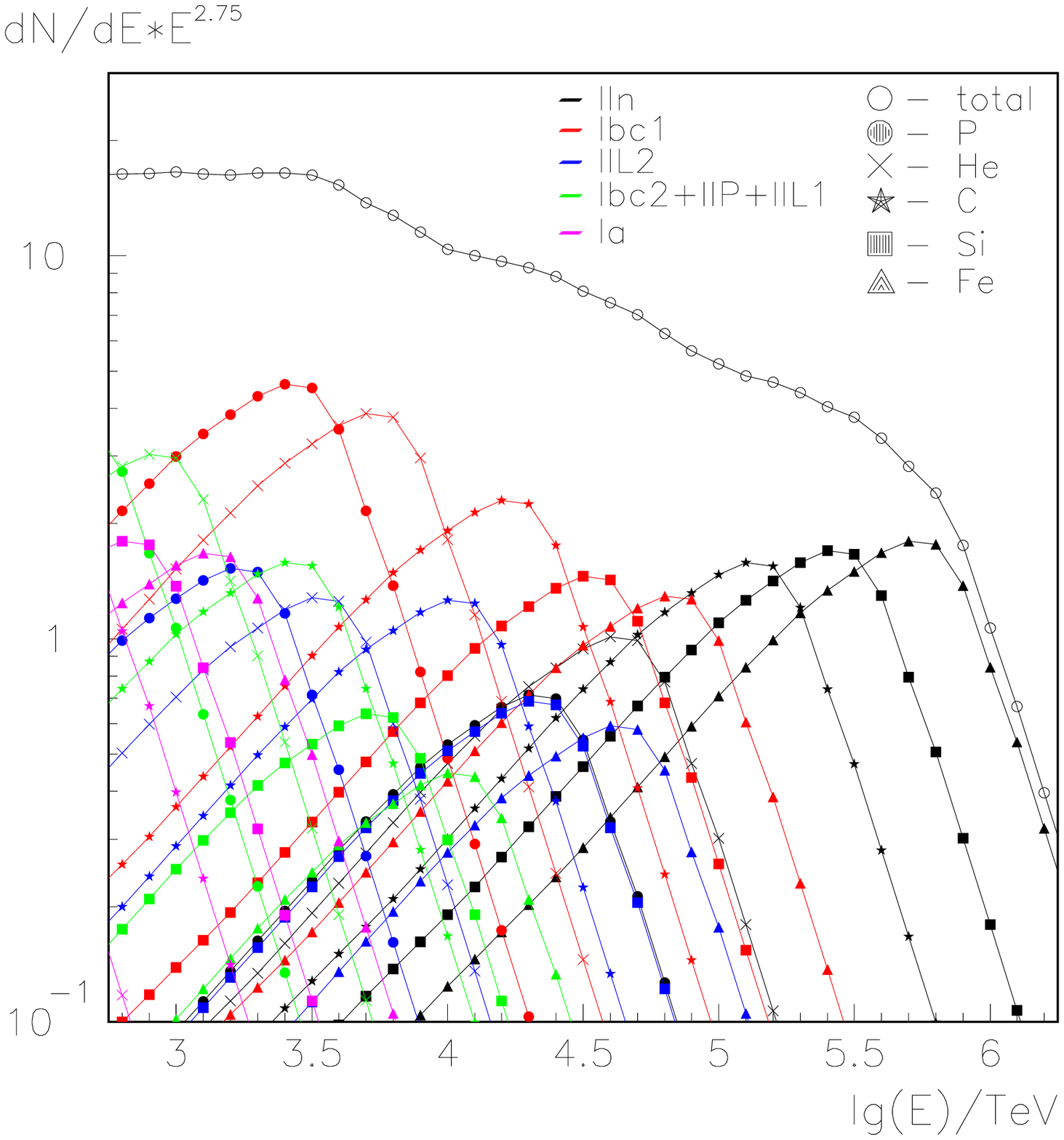}
\caption{\label{fig3} Left: Mean logarithmic mass from superposition of two CR sources. 
Right: Simulated spectra (in relative units) from SNe of different types.
}
\end{center}
\end{figure}

\section{Simulation of energy spectra}

Detailed observations of Supernovae show that nature of the phenomenon is complex, i.e. they don't present uniform class objects \cite{turatto}-\cite{hamuy}. 

First SNe classification was based on spectroscopic observations. 
Conventionally, 
all SNe were divided on two types. $I$-type comprises SNe with weak and unclear hydrogen lines in their spectra, whereas $II$-type is defined by bright hydrogen lines. 
In next stage these types were divided on several subtypes, which are differed in luminosities, expansion velocities, abundances, chemical composition of shell, etc.  

Some theoretical classifications give more importance to the origin of the explosion. They distinguish two fundamentally different SN types regardless of their spectroscopic appearance: core collapse and thermonuclear SNe. 
SNe II, Ib, Ic occur near star forming regions and have never been observed in elliptical galaxies. Their progenitors are massive stars born with more than 8 $M_\odot$ that undergo core collapse, leaving a neutron star or black hole as a remnant. Ia SNe are observed in all types galaxies. They arise from white dwarfs that explode as they approach the Chandrasekhar mass ($\sim 1.4 M_\odot$) after a period of mass accretion from binary companion, leaving no compact remnant behind them.

The present classification scheme is complex.

At that, only some subclasses of SNe can provide the knee particles while most SNe have spectra cutting off at considerably lower energies. 
In order to generate the knee not only the cut-off energy $E_c$ needs to be tuned, but also the relative flux intensities of the sources need to be adjusted. 
 
Simulation of elemental groups spectra is based on data of \cite{7a}. According to this work main deposit to total CR intensity in knee region are given by SNe Ibc1 and IIn types. SNe of IIL1, Ibc2 and IIP types have close $E_{c}$ and so they can bring large summarized deposit too. SNe Ia type have lower value of $E_{c}$ and so they can influence in knee region by heavy elements only. 
  
Fig.3right shows simulated deposit from SNe of different types. At that, a behavior of the experimental all-particle energy spectrum (before knee $\gamma \sim -2.75$, after knee $\gamma \sim -3.1$) was main criterion at choice of the mass composition of the sources. 

In Fig.4left it is shown summarized elemental group spectra from superposition of these Supernovae. 

As it is seen from the figure the knee is caused by the decreasing flux of light primaries, corroborating results of many astrophysical assumptions explaining the origin of the knee \cite{horandel}. 

\begin{figure}[t]
\begin{center}
\includegraphics*[width=0.48\textwidth,angle=0,clip]{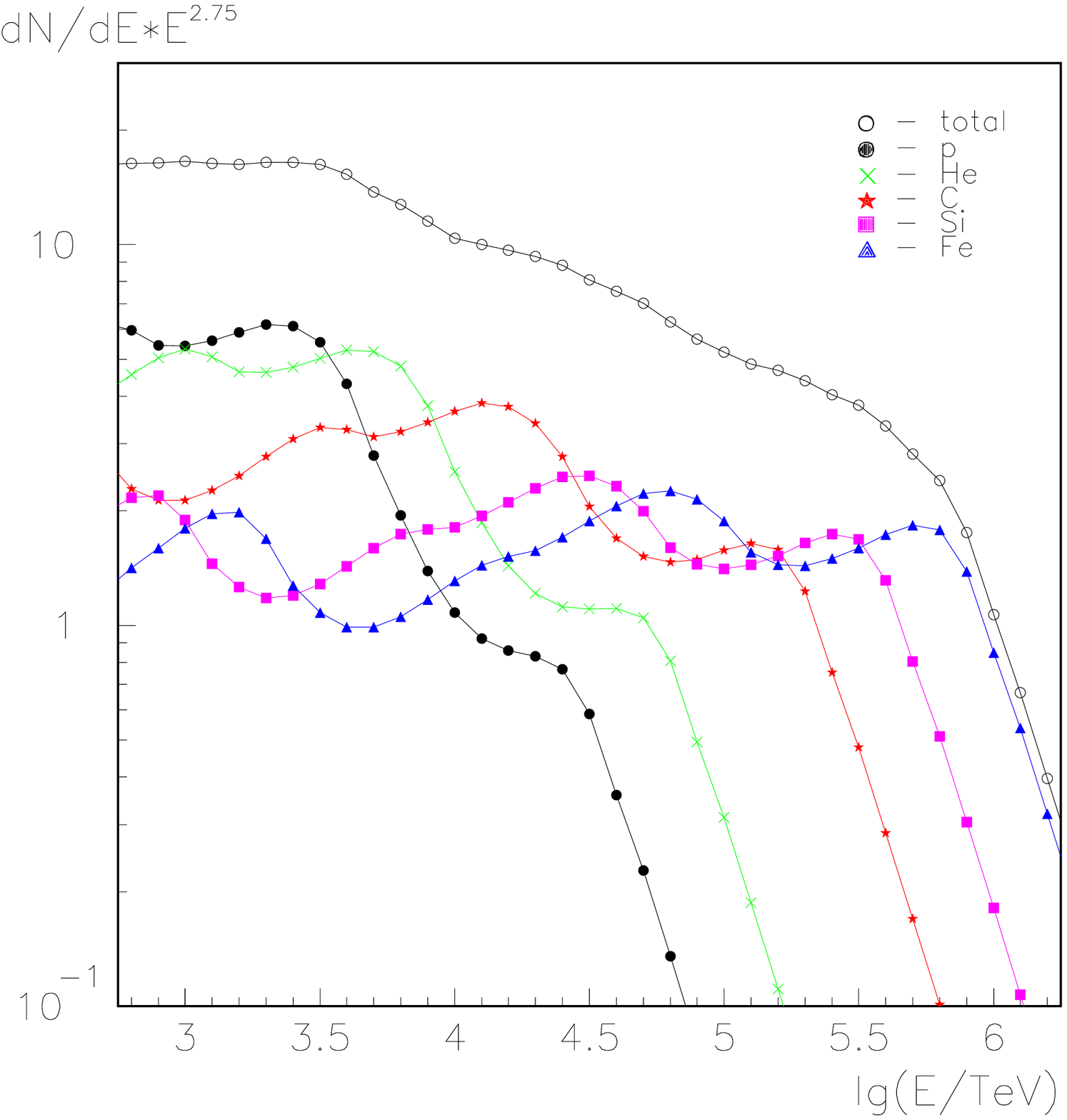}
\includegraphics*[width=0.48\textwidth,angle=0,clip]{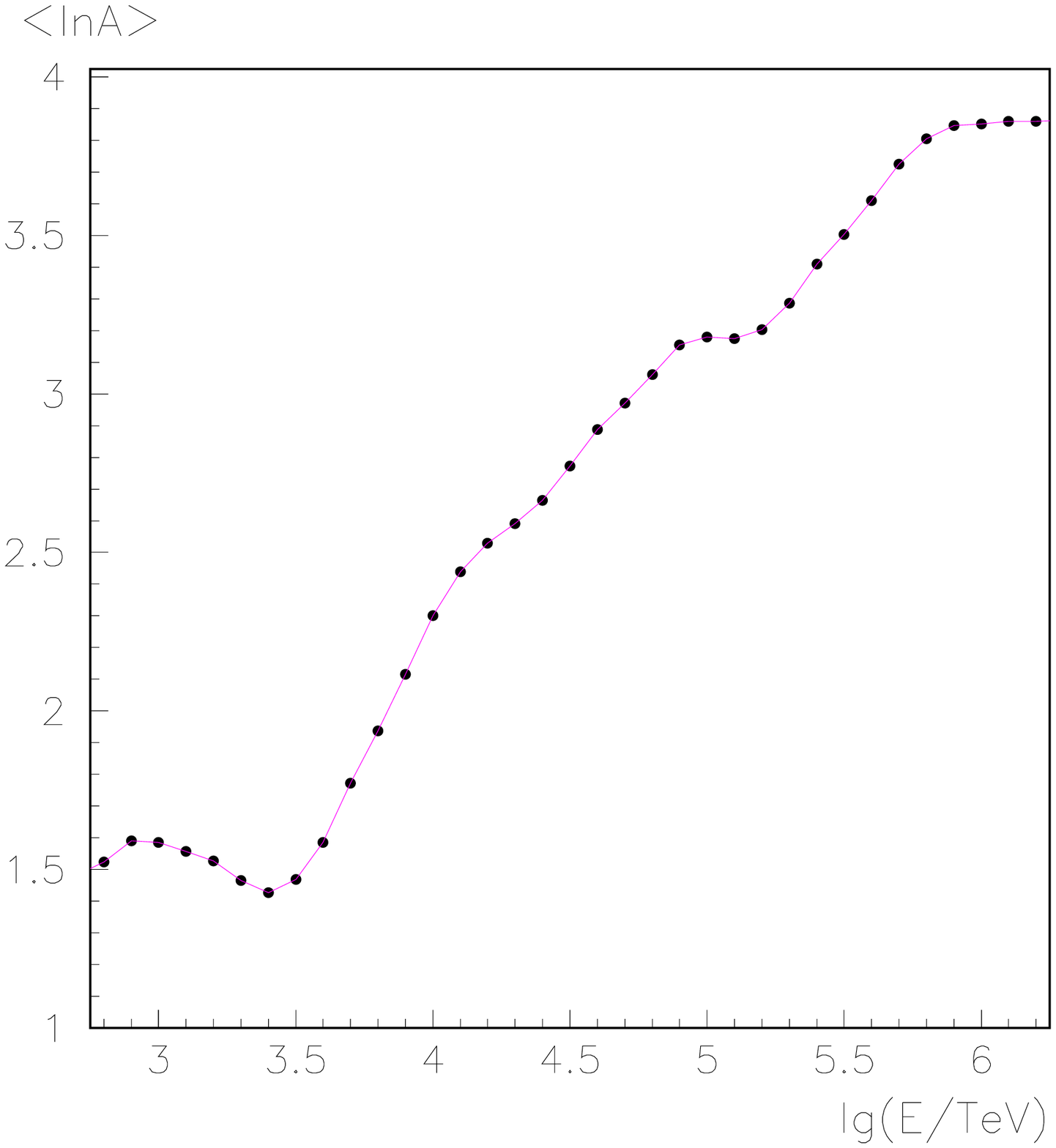}
\caption{\label{fig4} Left: Summarized spectra from superposition of SNe of different types. 
Right: Mean logarithmic mass vs. primary energy from superposition of SNe of different types. 
}
\end{center}
\end{figure}

The most important consequence of the presented analysis is a significant increase of the slope index in the energy spectrum of heavy element groups at energy of a few PeV. This effect is initiated by large difference between $E_{c}$ of SNe Ia type and neighboring SNe of IIL1, Ibc2 and IIP types.     

Fig.4right shows the resulting dependence of mean logarithmic mass on primary energy. From the Figure it is seen that the composition becomes heavier after the knee. The result is compatible with the evidence from the most of experiments \cite{yodh}. 

 
\section{Conclusion}

Spectra of 5 elemental groups (p, He, C, Si, Fe) are simulated on the base of a superposition of few CR sources (Supernovae of different types) with defined mass composition. The knee in all-particle energy spectrum is explained by decreasing flux of the light primaries. At that, the slopes of summarizing energy spectra for elemental mass groups of cosmic rays have significant changes. In particular, the significant increase of heavy elements spectrum slope at energy of few PeV, is shown.

\end{document}